\documentclass{article}

\usepackage{arxiv}

\usepackage[utf8]{inputenc} 
\usepackage[T1]{fontenc}    
\usepackage{hyperref}       
\usepackage{url}            
\usepackage{booktabs}       
\usepackage{amsfonts}       
\usepackage{nicefrac}       
\usepackage{microtype}      
\usepackage{lipsum}
\usepackage{graphicx}
\graphicspath{ {./images/} }
\usepackage{mathptmx}
\usepackage{cite}
\usepackage{amsmath,amssymb,amsfonts}
\usepackage{algorithmic}
\usepackage{graphicx}
\usepackage{textcomp}
\usepackage{xcolor}

\usepackage{graphicx}
\usepackage{subcaption}

\title{Alzheimer's Disease Classification using Retinal OCT: TransNetOCT and Swin Transformer Models}

\author{
 Siva Manohar Reddy Kesu \\
  Research Division\\
  AIT Resource Group Inc\\
  Trichy, India 620001 \\
  \texttt{smrkesu@aitrg.com} \\
   \And
 Neelam Sinha\\
  Centre for Brain Research\\
  Indian Institute of Science\\
  Bangalore, India 560012 \\
  \texttt{neelam@cbr-iisc.ac.in} \\
  \And
 Hariharan Ramasangu \\
  Research Division\\
  Relecura. Inc.\\
  Bangalore, India  560075 \\
  \texttt{rharihar@ieee.org} \\
   \And
 Thomas Gregor Issac\\
  Centre for Brain Research\\
  Indian Institute of Science\\
  Bangalore, India 560012 \\
  \texttt{thomasgregor@cbr-iisc.ac.in} \\
}

\begin{document}
\maketitle
\begin{abstract}
Retinal optical coherence tomography (OCT) images are the biomarkers for neurodegenerative diseases, which are rising in prevalence. Early detection of Alzheimer's disease using retinal OCT is a primary challenging task. This work utilizes advanced deep-learning techniques to classify retinal OCT images of subjects with Alzheimer's disease (AD) and healthy controls (CO). The goal is to enhance diagnostic capabilities through efficient image analysis. In the proposed model, Raw OCT images have been preprocessed with ImageJ and given to various deep-learning models to evaluate the accuracy. The best classification architecture is TransNetOCT, which has an average accuracy of 98.18\% for input OCT images and 98.91\% for segmented OCT images for five-fold cross-validation compared to other models, and the Swin Transformer model has achieved an accuracy of 93.54\%. The evaluation accuracy metric demonstrated TransNetOCT and Swin transformer models' capability to classify AD and CO subjects reliably, contributing to the potential for improved diagnostic processes in clinical settings.

\end{abstract}


\section{Introduction}

Alzheimer's disease (AD) is the most common form of dementia.  It is characterized by progressive cognitive impairment, memory deficit, and a decline in learning and executive functioning clinically.  AD is neuropathologically characterized by amyloid-beta (Ab)-plaques and neurofibrillary tangles containing tau.  These neuropathological changes are believed to develop 15–20 years before symptom onset.  AD is diagnosed in subjects with MCI or dementia using clinical criteria combined with abnormal biomarkers for Ab pathology or neuronal injury.  Ab pathology is reflected by decreased Ab levels in cerebrospinal fluid (CSF) or on an amyloid positron emission tomography (PET).  Neuronal injury is reflected by either cortical atrophy on magnetic resonance imaging (MRI), hypometabolism on fluorodeoxyglucose-PET (FDG-PET), or increased tau and phosphorylated tau (pTau) levels in CSF.  These biomarkers, however, are invasive, expensive, or time-consuming.  Thus, there is a requirement for an early, patient-friendly, inexpensive AD biomarker that preferably detects AD pathology before severe neurodegeneration \cite{ottoy2019association}.  There is increasing evidence that both anterior and posterior visual pathways are also affected in AD, which could, in turn, be responsible for the visual symptoms.  Several AD patients also display impairment in spatial contrast sensitivity, among other visual deficits.  The poor visual memory performance may be an early indication of AD, years before the actual diagnosis \cite{czako2020retinal} \cite{rehan2021visual}. 

The retina is the only neural tissue that can be directly visualized non-invasively in the human body.  Findings from retinal imaging can be informative regarding the brain's health; many abnormalities in retinal imaging have been linked with cerebral pathology \cite{bahr2024deep}.  The retina is embryologically obtained from the neural tube cranial part,, similar to the brain. The retina is easily accessible, and retinal neurons can be visualized through high-resolution optical methods such as optical coherence tomography (OCT), visualizing the thickness of retinal layers. With OCT, retinal changes are visualized both in ophthalmological and neurodegenerative diseases \cite{chauhan2020differential} \cite{ghita2023ganglion}.

 In a mouse model of AD, Ab deposits have been found in the retina, specifically in retinal ganglion cells (RGCs), consistent with the pathology observed in the brain.  Other labs have shown amyloid deposits and neurofibrillary tangles in AD patients' visual association area and subcortical visual centers.  A non-invasive method called OCT is used to observe abnormalities in the retinal structure in vivo in patients with AD.  Their work demonstrated a significant reduction in the overall thickness of the RNFL and within each quadrant in AD patients.  Other labs have reported similar findings, and histopathological changes in the eye have also been reported, such as degeneration of the optic nerve and loss of RGCs in AD patients.  The loss of optic fiber nerves has been predominantly due to a loss of M-cells (the largest class of RGCs).  The eye can diagnose and evaluate pathological changes in the brain in a non-invasive manner \cite{coppola2020optical} \cite{mirzaei2020alzheimer}.

Multiple postmortem pathology studies have identified beta-amyloid plaques and neurofibrillary tangles in the retina of patients with varying stages of Alzheimer's disease (AD). In patients with Parkinson's disease (PD), pathology studies have found lower levels of dopamine in the retina. The neurodegeneration that accompanies Huntington's disease (HD) and amyotrophic lateral sclerosis (ALS) may appear in the retina as well. Case-control comparisons suggest that patients with mild cognitive impairment (MCI) or unspecified dementia (D-US) also exhibit retinal thinning. Previous studies have shown that the RNFL and GCL thicknesses are reduced in subjects with multiple sclerosis (MS), Parkinson's disease (PD), and AD \cite{mirzaei2020alzheimer} \cite{zhang2022retina} \cite{hart2024neuropathological}.

MCI is a preclinical or early phase of AD. More than half the patients diagnosed with MCI eventually develop AD. The retinas in MCI patients and age-matched controls using OCT showed that the former had a comparatively thinner RNFL. MCI is generally defined as memory impairment, and it can be further divided into various subtypes, such as vascular, metabolic amnestic, etc. Among them, the amnestic subtype is considered to be a preclinical stage of AD. The study was focused on a-MCI patients suffering from multiple domains and having a significant problem of memory impairment. The anatomical abnormalities in the peripapillary RNFL and macula lutea in Alzheimer's disease and a-MCI patients to choose whether these changes have to be correlated with the severity of dementia in the patients has been investigated \cite{gao2015abnormal}.

The superior pRNFL is the best biomarker for identifying most AD cases. However, it could be more efficient when classifying between mild AD and MCI. The global pRNFL(pRNFL-G) is another reliable biomarker to discriminate frontotemporal dementia from mild AD and healthy controls (HCs), moderate AD, and MCI from HCs. Conversely, pRNFL-G fails to realize mild AD and the progression of AD. The average pRNFL thickness variation is considered a viable biomarker for monitoring AD progression. Finally, the superior and average pRNFL thicknesses are considered consistent for advanced AD but not for early/mild AD \cite{ibrahim2023systematic}.

A study has been performed meta-analysis to assess the retinal layer thickness in AD and MCI patients and cognitively normal subjects. They evaluated the role of concomitant ophthalmological disease on retinal thickness, particularly glaucoma, and the possible confounding role of age and disease severity \cite{den2017retinal}.

OCT, OCT angiography (OCT-A), and fundus imaging allow for detailed quantitative and qualitative analysis of retinal features. OCT uses the reflectivity of light to microimage the anatomy of the retina and optic disk. The pRNFL, and macular ganglion cell layer and inner plexiform layer (mGCIPL) are significantly implicated in neurodegenerative states. In contrast, other markers, such as macular volume and choroidal thickness, have also been studied. OCT-A compares retinal layers across time as blood flows through the capillaries. O T-A captures information regarding retinal vasculature, including microvascular density, branching complexity, and flow density. Fundus imaging allows for directly visualizing the macula, optic disk, and retinal vasculature. Vessel tortuosity and branching complexity have been identified as helpful biomarkers, and other retinal features can be directly visualized through fluorescence imaging. Each imaging modality provides a host of information that has revealed retinal manifestations of neurodegenerative disease \cite{bahr2024deep}.

Spectral-domain optical coherence tomography (SD-OCT) is an accessible clinical tool for measuring structural changes to the retina and increasingly as a biomarker for brain-predominant neurodegenerative diseases like Alzheimer's. Inspired by methods from other functional imaging modalities, acquired images while repeatedly cycling lights on and off and spatially normalized retinas to facilitate intra- and inter-individual analyses. The experiments used a relatively brief and easy-to-implement protocol to replicate and expand upon previous functional OCT findings in healthy adults, uncovering a potential new biomarker for Alzheimer's disease. Due to their spatial distribution and the impact of anti-AQ4 antibodies, it concluded that at least some light-dependent changes in retinal reflectivity report on glial cell function. This represents a non-invasive approach to studying glial function, which may be necessary for the pathobiology of several other neurodegenerative diseases \cite{bissig2020optical}.

 Machine learning algorithms assist us in detecting information in retinal images that can only be readily apparent with computational algorithms. Several studies have been conducted to determine if systemic patient health information can be collected from retinal images. Such algorithms have demonstrated good accuracy at predicting quantitative variables, such as coronary artery calcium or serum creatinine, and qualitative variables, using retinal fundus images alone. It is challenging to identify which features in the retinal images are used by the machine learning algorithms to glean the information. However, more information may be contained in retinal images than was previously known, and it is necessary to apply computational algorithms, such as machine learning, to establish many classic pathologic features of neurodegenerative disease \cite{kashani2021past} \cite{marchesi2021ocular}.

 Given the known correlation between retinal health and neurodegenerative disease, there is good potential for deep learning algorithms to ascertain information regarding cerebral disease from retinal images. Indeed, a growing amount of literature has documented correlations between the progression of neurodegenerative disease and physician-observable retinal findings, such as retinal arteriolar and venular caliber, vessel tortuosity, retinal layer thickness, and optic disc morphology. The research determines what information is contained within OCT, OCT-A, and color fundus images and what information cannot be obtained from retinal imaging \cite{turkan2024automated}.

A diagnostic tool for classifying MCI and AD using feature-based machine learning was applied to OCT-A. Multiple features are extracted from the OCT-A image, such as vessel density, foveal avascular zone area, retinal thickness, and novel features based on the histogram of the range-filtered OCT-A image. An extensive local database for our study was collected to ensure effectiveness for a diverse population. The promising accuracy of 92.17\%, has been achieved for the early diagnosis of AD \cite{visitsattapongse2024feature}.

The initiation for evaluating the retina, especially the retinal vasculature, is another approach for screening dementia patients using advanced machine-learning techniques that have been employed for early detection. Utilizing data from the UK Biobank, the pipeline achieved an average classification accuracy of 82.44\%. Besides the high classification accuracy, they added a saliency analysis to strengthen this pipeline's interpretability. The saliency analysis indicated that within retinal images, small vessels carry more information for diagnosing Alzheimer's diseases, which aligns with related studies \cite{tian2021modular}.

A bilateral model has been developed to detect Alzheimer's disease-dementia from retinal photographs. The EfficientNetb2 network has been used as the model to extract features from the images. The dataset consists of 12949 retinal photographs from 648 AD patients and 3240 healthy people used in the deep learning model training, validation, and test sets. The validation dataset from the deep learning model has achieved 83·6\% accuracy and area under the receiver operating characteristic curve (AUROC) of 0·93 (0·01). The bilateral deep learning mo el has achieved 92·1\% accuracy, and AUROC is 0·91 from the test dataset \cite{cheung2022deep}.

The multimodal CNN has been used on retinal images to successfully predict the diagnosis of symptomatic AD in an independent test set. The dataset with 222 e es of 123 cognitively normal subjects and 62 eyes of 36 AD subjects has been used to develop the model. AUROC of 0.841 for AD prediction from the test set consisti g of GC- IPL maps, quantitative data, and patient data \cite{wisely2022convolutional}.

The CNN models have been identified as early MCI diagnoses using datasets comprised of OCT and OCTA images. Unlike the OCTA images, the GC-IPL thickness has been mapped for decision support. The OCTA image dataset comprises 236 eyes with 129 cognitively healthy subjects and 154 eyes with 80 MCI subjects that have been used for the CNN model. The GC-IPL thickness maps, OCTA images, and quantitative OCT and OCTA data, the AUC value for the CNN was 0.809 when applied to the independent test set and achieved a sensitivity of 79\% and specificity of 83\%. Models using quantitative data alone were also tested, with a model using quantitative data derived from images, 0.960, outperforming a model using demographic data alone, 0.580  \cite{wisely2024convolutional}.

Another study found retinal OCT images to detect other diseases besides neurodegenerative ones. Retinal illnesses damage any portion of the retina, causing visual problems; some can even lead to blindness. Various retinal illnesses include Diabetic retinopathy, Macular pucker, Glaucoma, Macular hole, Age-related macular degeneration, Drusen, Central serous retinopathy, Macular edema, Vitreous traction, and Optic nerve abnormalities. The deep learning image identification system based on CNN is used to classify retinal illnesses precisely in their early stages. The OCT images of the retina are classified into "Age-related Macular Degeneration (AMD)  Choroidal Neovascularization (CNV), DRUSEN, Diabetic Macular Retinopathy (DMR), Diabetic Retinopathy (DR), Macular Hole (MH), Central Serous Retinopathy (CSR), and Normal Eyes" using a lightweight Deep neural network. The classification accuracy achieved in this study with the aid of VGG16 is around 97\% on the OCT images dataset \cite{subramanian2022classification}.

In this paper, the proposed work uses the retinal OCT Dryad dataset to classify AD and CO subjects. Various deep-learning architectures have been evaluated for classification accuracy, and transformer models have also been used.

\section{Proposed Methodology}

The increasing occurrence of neurodegenerative diseases such as AD has implied the development of robust diagnostic tools for early detection. This proposed work outlines the implementation of a retinal OCT image classification system using various deep-learning models to distinguish between Alzheimer's disease (AD) and healthy control (CO) subjects. This proposed work aims to improve diagnostic accuracy and contribute to medicine by leveraging advanced deep-learning and transformer techniques.

\subsection{Data Preparation}
 
The Dryad dataset \cite{bissig2020practical} consists of images categorized into AD and CO. Each class is organized into separate folders, with images stored in subdirectories for individual subjects. A Dryad dataset class was created to load and preprocess these images, applying necessary transformations, including resizing, normalization, and tensor conversion. The Dryad OCT dataset utilized in this study consists of images organized into two main folders representing AD classes: Images of subjects diagnosed with Alzheimer's disease. CO: Images of control subjects without neurodegenerative conditions have been depicted in Table \ref{tab:dataset}. The dataset is intended to facilitate classifying neurodegenerative conditions through deep learning methodologies.

OCT scans have been performed on a Heidelberg Engineering SPECTRALIS OCT. The SPECTRALIS Slice Planning Tool has been used to select each retina slice per subject, which is automatically re-selected for each acquisition. The single slice has been spanned 30° eccentricity and angled through the (para)fovea, the optic disc centre, generating the cross-sectional view. Automatic Real-time (ART) averaging has been set to a maximum of 100.  OCT images have been collected at after > 20 seconds of exposure to light ("LIGHT") or after 2 minutes of darkness ("DARK"), and five slices were taken in each mode. 
  
\begin{table}[h!]
\caption{Retinal OCT image Dataset}
\centering
\begin{tabular}{c c c c c}
 \toprule
\textbf{Sl No} & \textbf{Dataset Class} & \textbf{No of Subjects} & \textbf{Light Modes} & \textbf{Dark Modes}  \\ 
\midrule
1 & Alzheimer’s Disease & 14 & 5 images & 5 images \\ 
2 & Controlled & 14 & 5 images & 5 images \\ 
 \bottomrule
\end{tabular}
\label{tab:dataset}
\end{table}

\begin{figure}[!ht]
    \centering
    \fbox{\includegraphics[width=\linewidth]{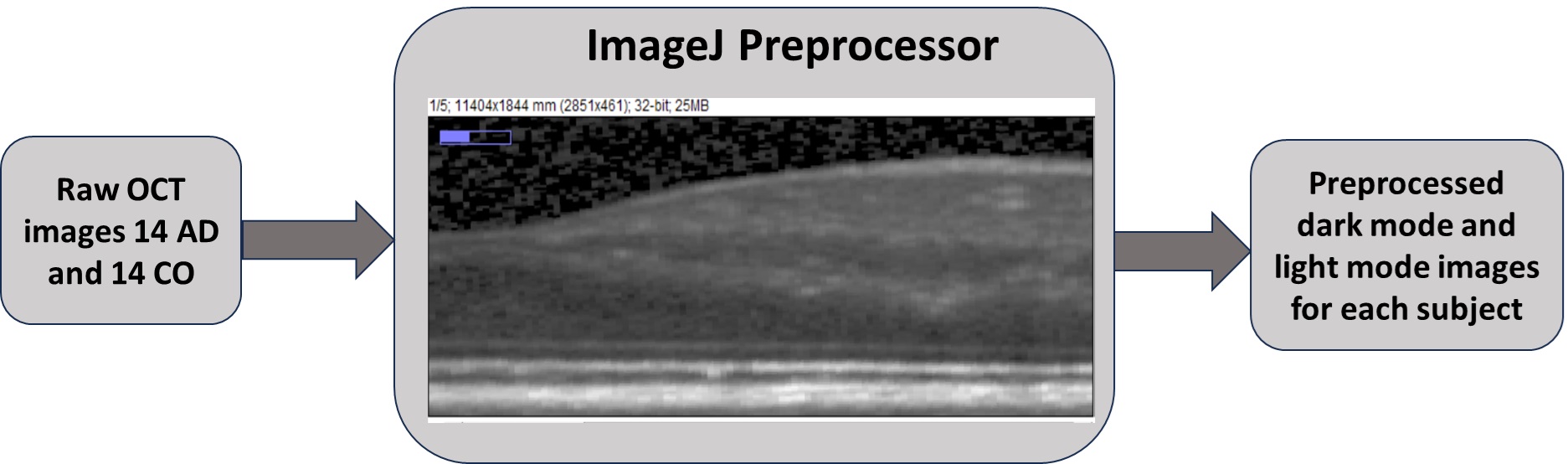} }
    \caption{Preprocessing raw OCT images}
    \label{fig:block_diagram1}
\end{figure}

OCT images are commonly preprocessed using ImageJ, an open-source image analysis tool widely used in medical imaging research. Preprocessing involves several steps to enhance the quality of OCT images for further analysis. Initially, noise reduction techniques such as median or Gaussian filtering are applied to minimize speckle noise, a common issue in OCT images. ImageJ's built-in filters help to smooth the OCT image without compromising essential retinal features. The brightness and contrast adjustments are performed to improve the visualization of retinal layers or specific areas of interest, ensuring crucial details for analysis. Raw Dryad datasets were extracted from sdb files using ImageJ shown in Figure \ref{fig:block_diagram1}. Retinas were cropped from raw reflectivity maps and resampled from the native image resolution to isometric pixels ($3.89 \mu m \times 3.89 \mu m$).

The data is structured so that each subject has a dedicated folder containing their respective image files in PNG or JPG format. A Dryad dataset class has been implemented to load and preprocess the images. The dataset has been initialized by loading image paths and corresponding labels. Images are resized to 224x224 pixels, a standard input size for different pre-trained models, and normalized image tensors are based on the mean and standard deviation of the ImageNet dataset. The OCT dataset was split into training and validation sets using an 80-20 split. 

\subsection{Model Architecture}

\begin{figure}[!ht]
    \centering
    \fbox{\includegraphics[width=\linewidth]{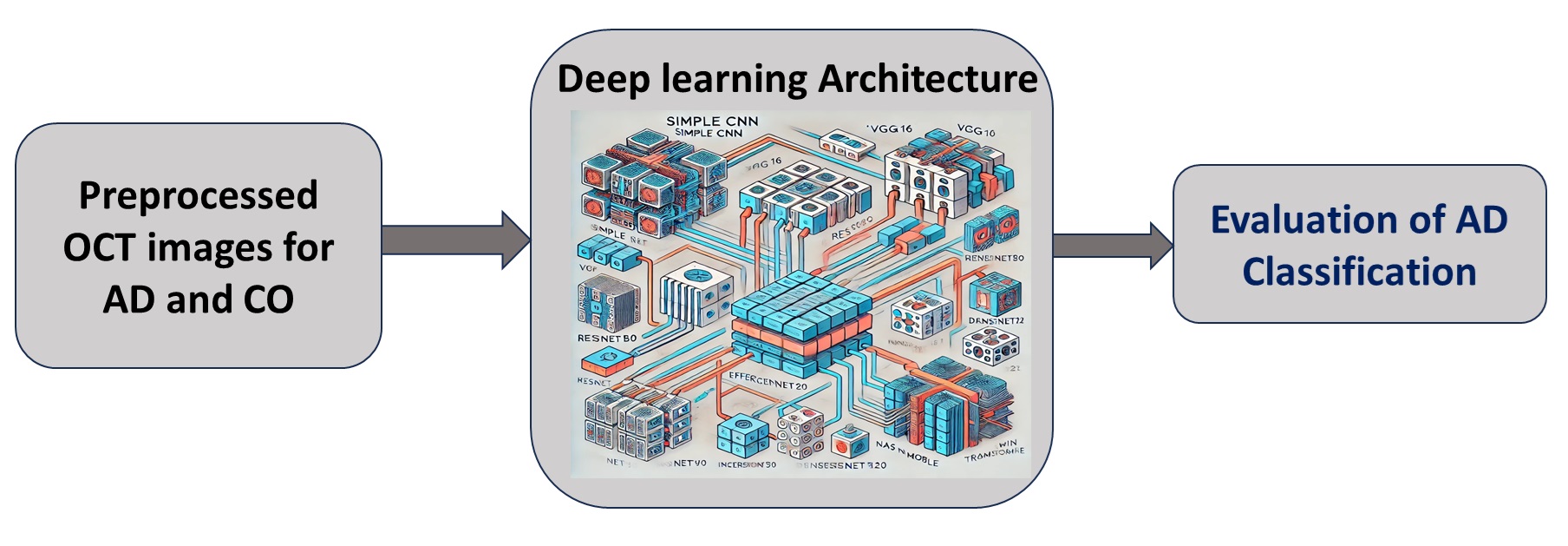} }
    \caption{Block diagram showcasing different deep learning architectures.}
    \label{fig:block_diagram2}
\end{figure}

The block diagram for the classification of the AD and CO OCT images using various deep learning architectures has been illustrated in Figure \ref{fig:block_diagram2}. The input to the deep learning architecture is the preprocessed images of 14 AD and 14 CO subjects OCT images. The evaluation accuracy of the classification of AD and CO has been performed using various deep learning architectures. Each block in figure \ref{fig:block_diagram2} corresponds to a different architecture. It illustrates the progression from simpler to more complex models and highlights the design components contributing to their varying performance levels.

A simple CNN is a basic architecture used in image classification and processing tasks. It typically consists of convolutional, pooling, and fully connected layers. These networks focus on extracting spatial hierarchies of features through convolutional filters. VGG16 is a more profound CNN architecture developed by the Visual Geometry Group (VGG). It consists of 16 layers with small 3x3 filters stacked on each other. This model is known for its simplicity in design and high performance in image classification tasks. RESNET50, part of the ResNet family, is a 50-layer deep neural network that introduced the concept of residual connections or "skip connections." These connections help solve the vanishing gradient problem that typically affects deep networks, allowing RESNET50 to maintain performance even with increasing depth \cite{mascarenhas2021comparison}. 

InceptionV3 is an improved version of the Inception architecture that uses various filter sizes and parallel convolution paths within each layer to capture multiple levels of abstraction \cite{jena2022convolutional}. EfficientNetB0 is part of the EfficientNet family, which scales the depth, width, and resolution of neural networks in a balanced manner. It aims to achieve high accuracy with fewer parameters compared to traditional architectures \cite{geetha2022classification}. DenseNet121 is a densely connected CNN architecture where each layer receives input from all preceding layers, promoting feature reuse and reducing the number of parameters. This helps in gradient propagation and improves learning efficiency \cite{arulananth2024classification}. NASNetMobile is a mobile-optimized version of the NASNet architecture, which was designed using Neural Architecture Search (NAS) to discover high-performing architectures \cite{sathishkumar2024detection} automatically. The Swin Transformer is a recent architecture based on the transformer model, initially developed for natural language processing tasks but adapted for vision tasks. It uses a hierarchical design and multi-head attention mechanisms to capture long-range dependencies in images \cite{liu2021swin}. 

A state-of-the-art Swin transformer model has been employed in this proposed work based on its performance in various image classification tasks. The model has been initialized using pre-trained weights, allowing it to leverage learned features from the ImageNet dataset, which enhances its ability to generalize to new image data. The Swin Transformer model has been chosen for its efficacy in image classification tasks. It has been initialized with pre-trained weights. This model architecture is designed for high performance in various computer vision tasks \cite{sun2023efficient}.

In addition, TransNet has been proposed for the OCT classification in this work. TransNet is a deep learning architecture that combines the benefits of transformers with other advanced methods, such as CNNs, to handle tasks related to image segmentation, disease diagnosis, or other medical imaging applications. Specifically, TransNet has gained attention for its application in segmenting the retinal layer from OCT scans \cite{elsharkawy2025transnetoct}. The Average Symmetric Surface Distance (ASSD) has been used to assess accuracy in OCT segmentation tasks, which measures the average bidirectional distance between the predicted segmentation surface and the ground truth surface. Combining ASSD with DSC and HD provides a balanced evaluation for OCT segmentation. DSC is an overlap agreement, and HD is sensitive to significant outliers. The mean absolute error (MAE) is used to calculate the average error for the thickness or boundaries of the layer. It is also suitable for OCT, where specific layer thickness is diagnostically necessary. Pixelwise Accuracy measures the percentage of correctly segmented pixels. Still, it can be less informative for thin structures like OCT layers, and the Boundary F1 Score evaluates precision and recall specifically at the boundaries \cite{koteimedical}.

\subsection{Training and Evaluation}

The dataset has been split into training and validation subsets. The training process involved optimizing the model using the cross-entropy loss function and the Adam optimizer  over multiple epochs. The model has been trained using cross-entropy loss as the criterion for measuring the model's performance. The Adam optimizer has been utilized to update the model weights during training. The training loop has been structured to iterate through the training dataset for a specified number of epochs, computing the loss and accuracy for each epoch. The model's performance has been evaluated on the validation dataset after training. Accuracy, precision, recall, and F1-score metrics have been calculated, and a confusion matrix has been plotted to visualize the classification performance. 

Class Activation Mapping (CAM) is a technique used to visualize which parts of an image are most relevant to a model's prediction. CAM generates heat maps highlighting important regions that contribute to the final decision, making it particularly useful for interpreting deep learning models. CAM works by using the weights of the final convolutional layer in a neural network and multiplying them with the feature maps to produce a localization map. This map helps identify the areas in an image most influential in predicting a specific class. CAM is valuable in medical imaging, where it provides interpretability to models like the Swin Transformer by showing which regions in an image, such as a tumor or lesion, were crucial in making a classification decision.

Five-fold cross-validation has been used to assess the performance and generalization of Swin Transformers model for retinal OCT image classification. The final performance is averaged across all five folds, providing an accurate and unbiased estimate of the model's effectiveness. This technique helps reduce overfitting and ensures the model is able to generalize well to unseen data, which is particularly important when training the Swin Transformer model on limited datasets.

The Swin Transformer model is a hierarchical vision transformer that significantly benefits from cross-validation of image classification and segmentation due to its complex architecture and multiple feature extraction stages. The Swin Transformer's self-attention mechanism learns from diverse subsets of the data, improving its ability to capture global and local features when applied with five-fold cross-validation. This approach helps fine-tune the model's hyperparameters and reduce the risk of overfitting, which is especially important given its capability to model long-range image dependencies. Cross-validation ensures the Swin Transformer performs consistently across different subsets of the data, providing insights into the model performance on new and unseen images.

The TransNetOCT model involves leveraging labeled OCT datasets to optimize the network's ability to perform segmentation and classification tasks. During training, combining transformers and CNNs, the hybrid architecture learns to extract meaningful spatial and contextual features from OCT scans. Data augmentation techniques, such as flipping, rotating, and normalizing the intensity, are applied to enhance generalization and robustness \cite{elsharkawy2025transnetoct}. The network is trained using loss functions tailored to the task, such as cross-entropy for classification and Dice loss for segmentation, and optimized with algorithms like Adam or SGD. Evaluation involves testing the trained model on a separate dataset to measure performance using accuracy, precision, recall, Dice coefficient, and area under the ROC curve (AUC). Attention maps generated during inference provide interpretability by highlighting regions crucial for decision-making, while robust validation ensures the network's effectiveness across diverse OCT imaging conditions and disease states.

\section{Results and Discussion}

The Dryad OCT dataset has been divided into 20\% testing and 80\% training data, run over ten epochs. The average loss has decreased over epochs, which shows that the models learned effectively from the training data. The model's accuracy has been calculated at the end of the epoch, which reflects the percentage of the correct predictions. The decrease in loss and increase in accuracy shows that the model effectiveness in learning. The evaluation metrics illustrate the models reliability in differentiating AD and CO subjects' OCT images, which provides the potential for improving diagnostic processes in clinical settings.

\begin{table}[h!]
\caption{Comparison of Deep Learning Architectures based on Accuracy}
\centering
\begin{tabular}{clc}
 \toprule
\textbf{Sl No} & \textbf{Deep Learning Architecture} & \textbf{Accuracy} \\ 
 \midrule
1 & Simple CNN & 61.818\% \\ 
2 & VGG16 & 67.27\% \\ 
3 & RESNET50 & 58.18\% \\ 
4 & InceptionV3 & 74.55\% \\ 
5 & EfficientNetB0 & 50.9\% \\ 
6 & DenseNet121 & 76.36\% \\ 
7 & NASNetMobile & 69.09\% \\ 
8 & Swin Transformer & 93.54\% \\ 
9 & TransNetOCT & 98.18\% \\ 
 \bottomrule
\end{tabular}

\label{tab:dl_architectures}
\end{table}

Table \ref{tab:dl_architectures} compares different deep learning architectures based on their accuracy performance in a given task. It lists eight architectures, including traditional CNN models like Simple CNN and more advanced ones like VGG16, RESNET50, and DenseNet121. The Simple CNN architecture has the lowest accuracy at 61.818\%, while VGG16 and RESNET50 show moderate performances with 67.27\% and 58.18\%, respectively. Other architectures, such as InceptionV3 and NASNetMobile, also show varying accuracies, with InceptionV3 performing relatively well at 74.55\%.

Table \ref{tab:dl_architectures} highlights that DenseNet121 and Swin Transformer perform significantly better than the rest, with DenseNet121 achieving 76.36\% accuracy and Swin Transformer reaching the highest accuracy of 98.11\%. EfficientNetB0, another well-known architecture, has a lower accuracy of 50.9\%, which is below the average of the architectures listed. This suggests that different architectures vary widely in their effectiveness, with Swin Transformer outperforming traditional CNN-based models.

\begin{figure}[!ht]
    \centering
    \fbox{\includegraphics[width=5cm, height=5cm]{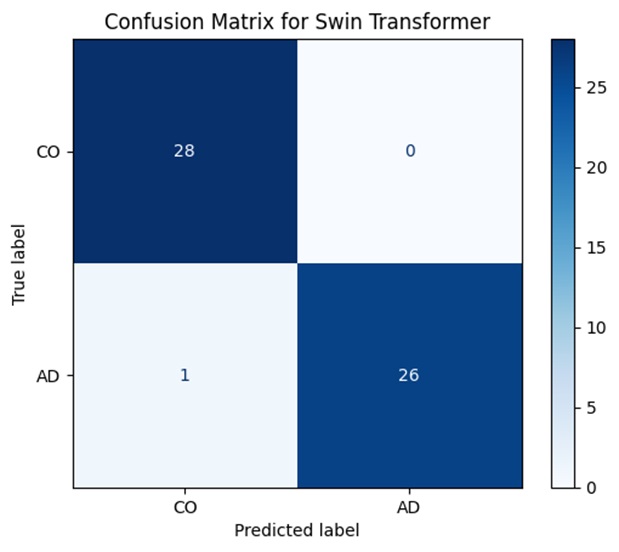} }
    \caption{Confusion Matrix for Swin Transformer model.}
    \label{fig:cm}
\end{figure}

Figure \ref{fig:cm} shows the confusion matrix for the test data from the Swin transformer deep learning model. The confusion matrix from the Swin Transformer model shows that out of 29 samples belonging to class CO, 28 have been correctly classified, while one has been misclassified as class AD. All 26 samples of class AD were correctly classified, with no false positives. The validation loss is 0.0575; the precision is 0.9524, which reflects the accuracy of positive predictions; the recall is 1.0000, which measures the model's ability to find all the true positive cases; and the F1 Score is 0.9811. The model demonstrates strong performance with very few misclassifications.

\begin{figure}[!ht]
    \centering
    \fbox{\includegraphics[width=8cm, height=6cm]{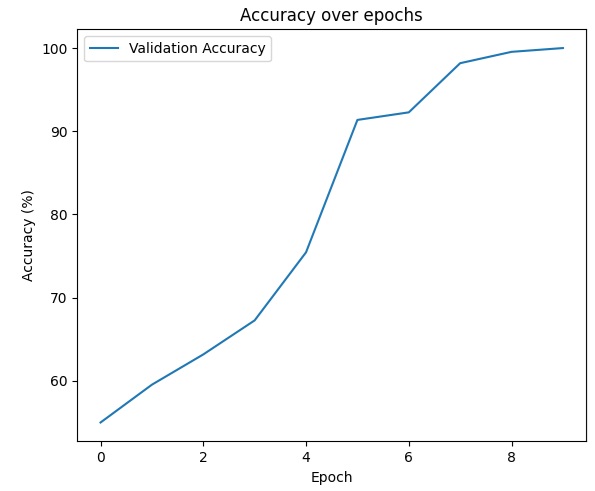}} 
    \caption{Accuracy over epochs}
    \label{fig:acc1}
\end{figure}

The validation accuracy per epoch has been shown in Figure \ref{fig:acc1}. The accuracy has reached near 100\% when it reaches the 10th epoch. At the 0th epoch, the model starts with a validation accuracy 50\%, indicating that it makes predictions only slightly better than random guessing. As training progresses, by the 10th epoch, the validation accuracy rises significantly to 98.11\%, showing that the model has learned and generalized well to the validation data. The enhancement of the accuracy over the ten epochs indicates that the transformer model explicitly captures the underscoring patterns in the dataset. The model has achieved a high level of performance in a relatively short number of epochs, showing efficient learning.

\begin{figure}[!ht]
    \centering
    \fbox{\includegraphics[width=10cm, height=5cm]{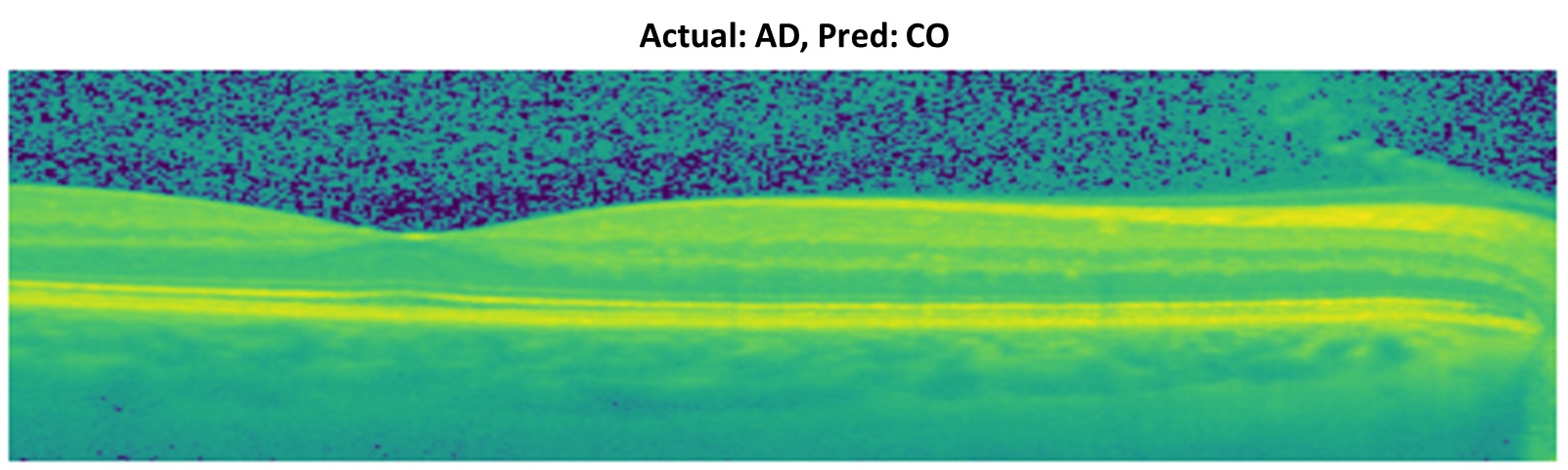}} 
    \caption{Misclassified image from test dataset}
    \label{fig:mci}
\end{figure}

Figure \ref{fig:mci} depicts the misclassified image from the test dataset. The OCT image with an actual label of AD is misclassified as CO, and the subject is AD613. The RNFL and choroid layers are essential in classifying Alzheimer's disease and healthy controlled subjects. Misclassification happens for several reasons, such as class similarity, where the model struggles to differentiate between Alzheimer's disease and healthy, controlled, visually similar categories. It also occurs due to underfitting, where the model has not learned enough distinguishing features, or due to imbalanced data, which biases the model toward predicting the majority class, leading to more false negatives for class AD. The misclassification affects the performance metrics, especially recall for class AD, and increases the false negative rate, where subjects of class AD are missed.

\subsection{Swin Transformer}

This proposed work detailed the implementation of an image classification model using a pre-trained Swin Transformer to distinguish between Alzheimer's disease and control subjects. The model effectively leveraged deep learning techniques to improve diagnostic accuracy. The five-fold cross-validation has been used to strengthen the robustness of the Swin transformer model further in this study. The Swin Transformer model achieves consistent accuracy across different data splits, depicting stability based on the mean accuracy across the folds. The average accuracy for the five-fold cross-validation is 93.45\%. Each fold in cross-validation serves as both a training and validation set at different stages, helping to verify the model's generalization to the unseen OCT dataset. Applying CAM interpretability with the Swin Transformer provides meaningful insights into the model's decision-making process for Alzheimer's disease. The CAM-generated heatmaps underline the regions of the OCT image that have been most predominant in the model's prediction.

\begin{figure}[!ht]
    \centering
    \begin{subfigure}[b]{0.45\textwidth}
        \centering
         \includegraphics[width=\linewidth]{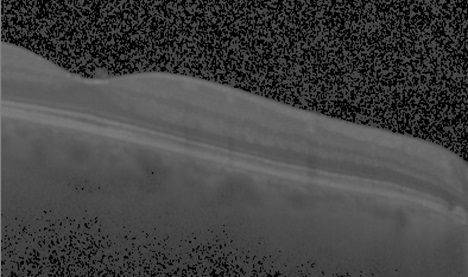}
        \caption{Input OCT image}
        \label{AD647a1}
    \end{subfigure}
    \begin{subfigure}[b]{0.47\textwidth}
        \centering
        \includegraphics[width=\textwidth]{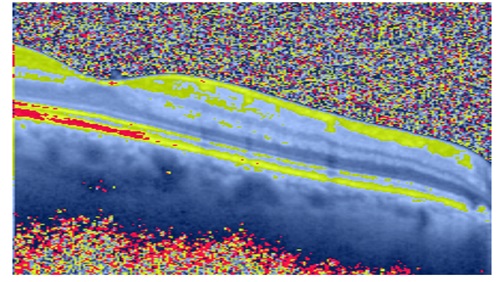}
        \caption{CAM OCT image}
        \label{AD647a2}
    \end{subfigure}

    \caption{Correctly classified Alzheimer's disease OCT image of AD647}
   \label{fig:ADCC1}
\end{figure}

Figure \ref{fig:ADCC1} contains two images showing an OCT retinal scan in dark mode for subject AD647 correctly classified by the model for detecting Alzheimer's disease. Figure \ref{AD647a1} appears to be a minimally processed OCT scan. The retina's layers are visible, with subtle gradients in grayscale indicating the boundaries between different retinal layers. Figure \ref{AD647a2} is a CAM heatmap overlaid on the OCT image, showing which areas the model focused on during classification. In this heatmap, warmer colors (like red and yellow) represent areas of high activation, indicating regions that contributed strongly to the model's decision. Cooler colors (like blue) represent areas with lower activation. The activation map focuses on specific bands within the retina, aligning with retinal layers that are the thinning of the RNFL and GCL, which are affected in Alzheimer's patients.

\begin{figure}[!ht]
    \centering
    \begin{subfigure}[b]{0.45\textwidth}
        \centering
         \includegraphics[width=\linewidth]{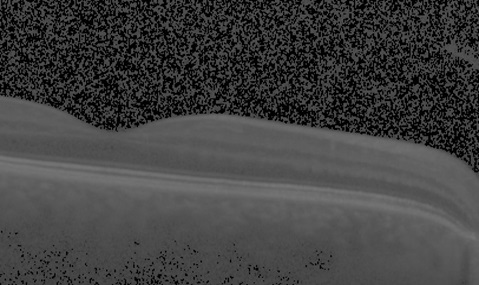}
        \caption{Input OCT image}
        \label{AD508a11}
    \end{subfigure}
    \begin{subfigure}[b]{0.45\textwidth}
        \centering
        \includegraphics[width=\textwidth]{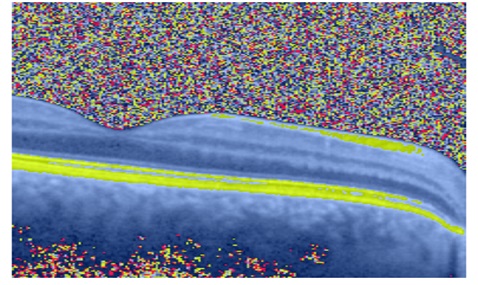}
        \caption{CAM OCT image}
        \label{AD508a21}
    \end{subfigure}

    \caption{Correctly classified Alzheimer's disease OCT image of AD508}
   \label{fig:ADCC2}
\end{figure}

Figure \ref{fig:ADCC2} depicts the correctly classified AD patient input retinal OCT image in dark mode and the corresponding CAM image for subject AD508. CAM in OCT imaging highlights essential areas in the retinal image for an Alzheimer's diagnosis by red and yellow color. The CAM highlights RNFL, choroid, and GCL areas, which are known to undergo thinning in Alzheimer's patients. Such changes are significant because they mirror neurodegeneration in the brain and can serve as non-invasive biomarkers for Alzheimer’s.

\begin{figure}[!ht]
    \centering
    \begin{subfigure}[b]{0.45\textwidth}
        \centering
         \includegraphics[width=\linewidth]{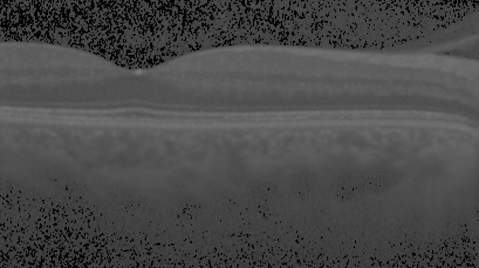} 
        \caption{Input OCT image}
        \label{CO578a1}
    \end{subfigure}
    \begin{subfigure}[b]{0.45\textwidth}
        \centering
        \includegraphics[width=\textwidth]{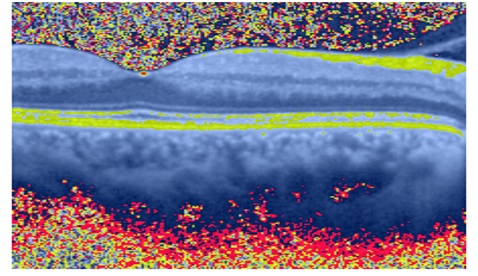} 
        \caption{CAM OCT image}
        \label{CO578a2}
    \end{subfigure}

    \caption{Correctly classified healthy controlled OCT image of CO578}
   \label{fig:COCC1}
\end{figure}

Figure \ref{fig:COCC1} depicts the correctly classified healthy patient input retinal OCT image in light mode and corresponding CAM image for subject CO578. After analyzing an OCT image, the CAM highlights regions of the retina that activate most strongly for classification as  Alzheimer 's-negative. The thick RNFL, GCL, and choroid areas show the absence of neurodegeneration.

\begin{figure}[!ht]
    \centering
    \begin{subfigure}[b]{0.45\textwidth}
        \centering
         \includegraphics[width=\linewidth]{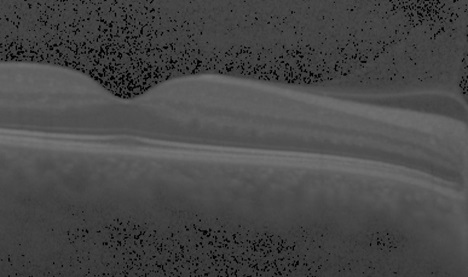} 
        \caption{Input OCT image}
        \label{CO942a1}
    \end{subfigure}
    \begin{subfigure}[b]{0.47\textwidth}
        \centering
        \includegraphics[width=\textwidth]{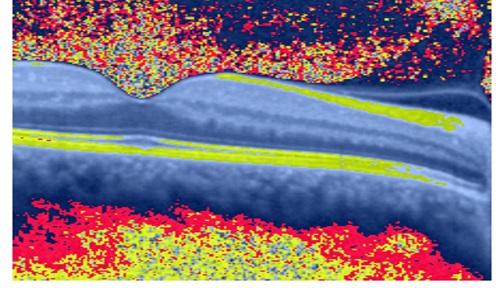}
        \caption{CAM OCT image}
        \label{CO942a2}
    \end{subfigure}

    \caption{Correctly classified healthy controlled OCT image of CO942}
   \label{fig:COCC2}
\end{figure}

Figure \ref{fig:COCC2} illustrates the correctly classified healthy patient input retinal OCT image in light mode and corresponding CAM image for subject CO942. CAM images focus on a thick RNFL layer, GCL layer, choroid, and optic nerve head associated with healthy control. 
The Swin Transformer model's focus on clinically relevant retina layers and the reduced influence of noise contribute to the accurate classification.


\begin{figure}[!ht]
    \centering
    \begin{subfigure}[b]{0.44\textwidth}
        \centering
         \includegraphics[width=\linewidth]{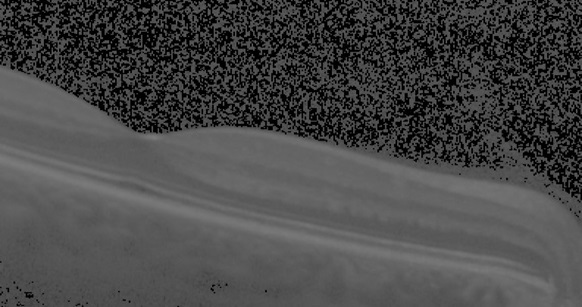}
        \caption{Input OCT image}
        \label{AD613a1}
    \end{subfigure}
    \begin{subfigure}[b]{0.47\textwidth}
        \centering
        \includegraphics[width=\textwidth]{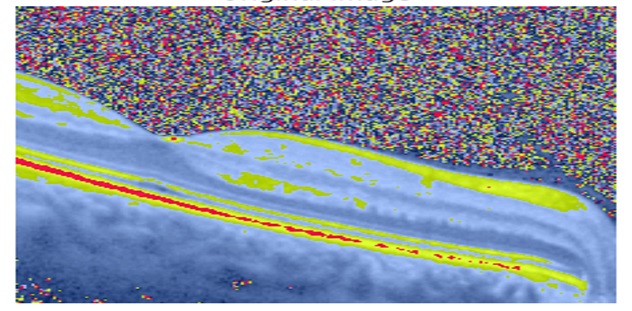}
        \caption{CAM OCT image}
        \label{AD613a2}
    \end{subfigure}

    \caption{Misclassified Alzheimer's disease OCT images of AD613}
    \label{fig:MISC1}
\end{figure}

Figure \ref{fig:MISC1} depicts the misclassified AD patient input retinal OCT image in light mode and the corresponding CAM image for subject AD613. CAM image shows that the RFNL and choroid layers are slightly thicker in the subject AD613, which has been misclassified.

\begin{figure}[!ht]
    \centering
    \begin{subfigure}[b]{0.45\textwidth}
        \centering
         \includegraphics[width=\linewidth]{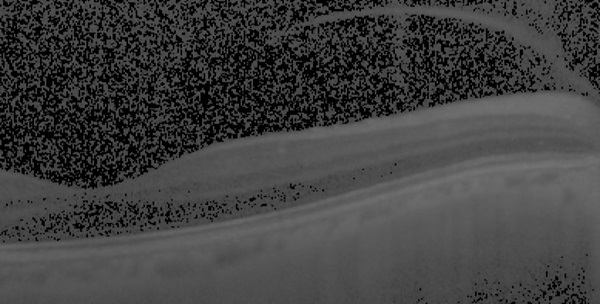} 
        \caption{Input OCT image}
        \label{CO127a1}
    \end{subfigure}
    \begin{subfigure}[b]{0.47\textwidth}
        \centering
        \includegraphics[width=\textwidth]{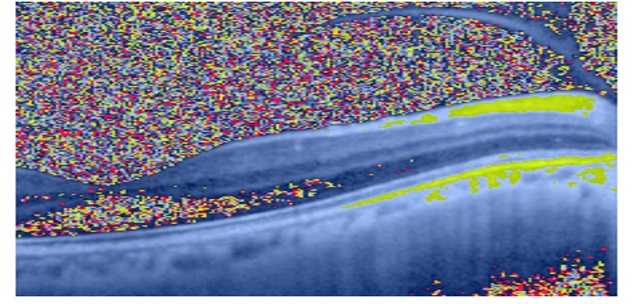} 
        \caption{CAM OCT image}
        \label{CO127a2}
    \end{subfigure}

    \caption{Misclassified healthy controlled OCT images of CO127}
    \label{fig:MISC2}
\end{figure}

Figure \ref{fig:MISC2} depicts the misclassified AD patient input retinal OCT image in light mode and the corresponding CAM image for subject CO127, which enhances interpretability in Alzheimer's diagnosis using OCT images. The thick RNFL layer region plays a vital role in the misclassification, which is healthy controlled but detected as Alzheimer's disease. 


Using CAMs with OCT for Alzheimer’s disease helps bridge retinal imaging and neurodegenerative diagnosis by identifying retinal biomarkers that serve as proxies for brain pathology, all through an accessible and less invasive imaging modality. The RNFL, choroid layer, and also sclera regions play a vital role in the classification of AD and CO subjects based on the Figures \ref{fig:ADCC1}, \ref{fig:ADCC2}, \ref{fig:COCC1}, \ref{fig:COCC2}, \ref{fig:MISC1}, and \ref{fig:MISC2}.

Compared to misclassified OCT images, the correctly classified OCT image appears less noisy, especially in RNFL, GCL, and choroid regions where the model has high activation. The model’s focus on specific retinal regions exhibiting Alzheimer ’s-related changes demonstrates that it captures features aligned with known disease biomarkers. This alignment is essential in clinical applications because it indicates that the model’s decision-making process is based on established medical knowledge, increasing confidence in its output.


The results of a five-fold cross-validation experiment with a Swin Transformer model applied to segmented OCT images, achieving an average accuracy of 89.82\%.

\begin{figure}[!ht]
    \centering
    \begin{subfigure}[b]{0.31\textwidth}
        \centering
         \includegraphics[width=\linewidth]{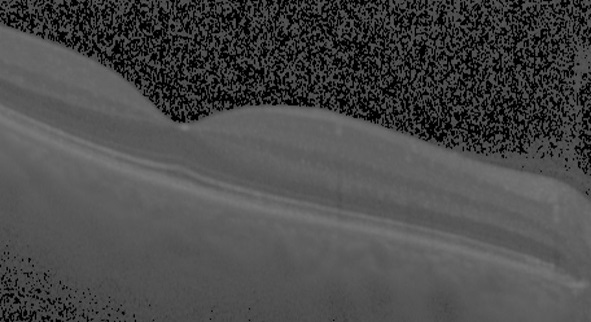}
        \caption{Input OCT image}
        \label{AD439a1}
    \end{subfigure}
    \begin{subfigure}[b]{0.29\textwidth}
        \centering
         \includegraphics[width=\linewidth]{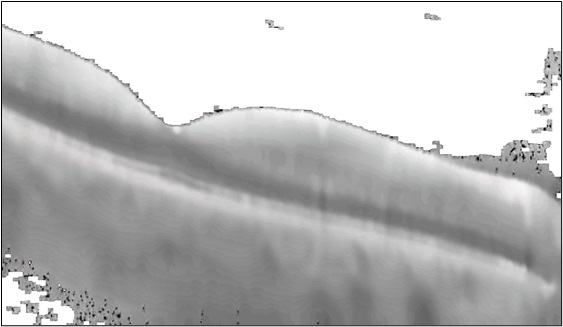}
        \caption{Segmented OCT image}
        \label{AD439a2}
    \end{subfigure}
    \begin{subfigure}[b]{0.29\textwidth}
        \centering
        \includegraphics[width=\textwidth]{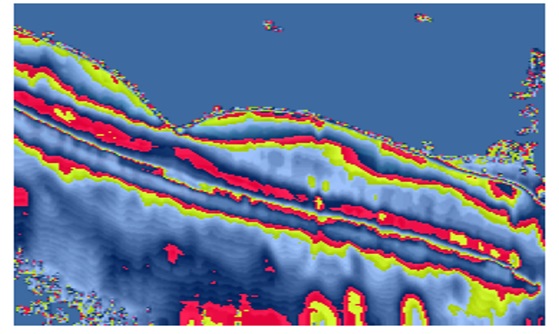} 
        \caption{CAM OCT image}
        \label{AD439a3}
    \end{subfigure}

    \caption{Correctly classified Alzheimer's disease segmented OCT images of AD439}
    \label{fig:ADSCC1}
\end{figure}

Figure \ref{fig:ADSCC1} depicts the correctly classified AD patient input OCT image, segmented retinal OCT image in dark mode, and the corresponding CAM image for subject AD439. Figure \ref{AD439a1} is a minimally processed OCT scan of the retina. The shades of gray represent various layers and structures in the retina, with darker regions indicating less reflective areas. Figure \ref{AD439a2} appears to be a binary or thresholded version of the OCT scan. In this representation, areas are segmented into black-and-white regions, possibly highlighting specific structures within the retina, particularly retinal layers. Figure \ref{AD439a3} is a CAM heatmap showing the areas the model considers essential for classification. Red and yellow indicate regions with high activation, meaning these areas strongly influenced the model’s decision. The blue color represents areas with low activation.


\begin{figure}[!ht]
    \centering
    \begin{subfigure}[b]{0.31\textwidth}
        \centering
         \includegraphics[width=\linewidth]{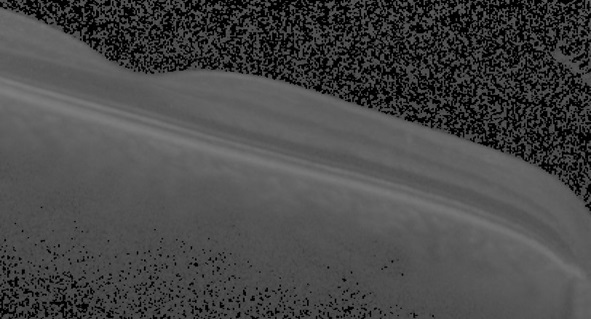}
        \caption{Input OCT image}
        \label{AD508a1}
    \end{subfigure}
    \begin{subfigure}[b]{0.31\textwidth}
        \centering
         \includegraphics[width=\linewidth]{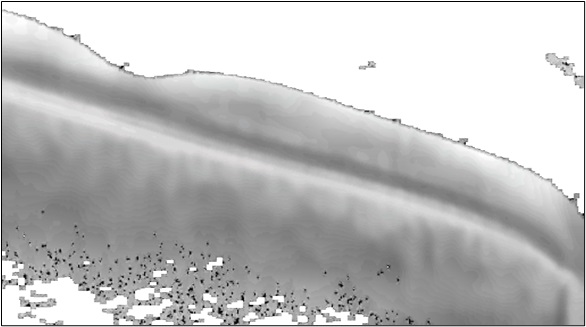}
        \caption{Segmented OCT image}
        \label{AD508a2}
    \end{subfigure}
    \begin{subfigure}[b]{0.30\textwidth}
        \centering
        \includegraphics[width=\textwidth]{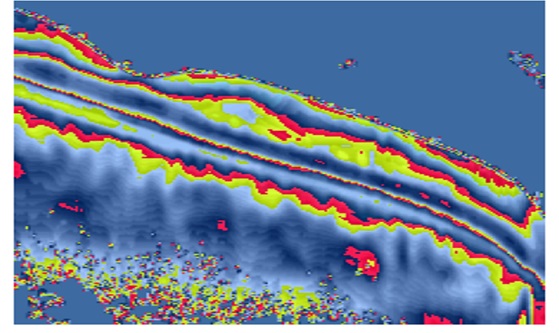}
        \caption{CAM OCT image}
        \label{AD508a3}
    \end{subfigure}

    \caption{Correctly classified Alzheimer's disease segmented OCT images of AD508}
    \label{fig:ADSCC2}
\end{figure}

Figure \ref{fig:ADSCC2} depicts the correctly classified AD patient input OCT image, segmented retinal OCT image in dark mode, and corresponding CAM image for subject AD508. The CAM shows that the model concentrates on the correct retinal layers related to Alzheimer’s pathology, such as the RNFL and GCL, which undergo thinning or other structural changes in Alzheimer’s patients. The targeted region focuses on relevant retinal structures, suggesting the model has learned to identify areas clinically linked to the disease, improving classification accuracy.

\begin{figure}[!ht]
    \centering
    \begin{subfigure}[b]{0.31\textwidth}
        \centering
         \includegraphics[width=\linewidth]{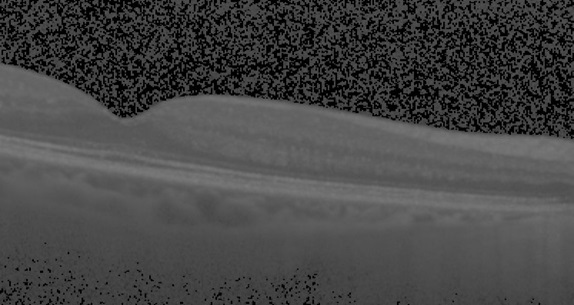}
        \caption{Input OCT image}
        \label{CO638a1}
    \end{subfigure}
    \begin{subfigure}[b]{0.31\textwidth}
        \centering
         \includegraphics[width=\linewidth]{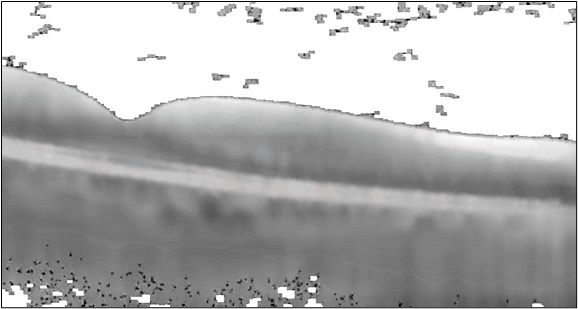}
        \caption{Segmented OCT image}
        \label{CO638a2}
    \end{subfigure}
    \begin{subfigure}[b]{0.30\textwidth}
        \centering
        \includegraphics[width=\textwidth]{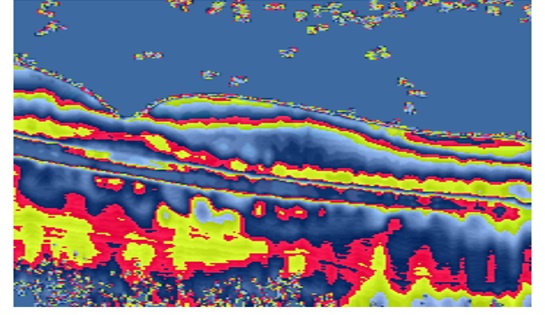}
        \caption{CAM OCT image}
        \label{CO638a3}
    \end{subfigure}

    \caption{Correctly classified healthy control segmented OCT images of subject CO638}
    \label{fig:COSCC1}
\end{figure}

Figure \ref{fig:COSCC1} depicts the correctly classified healthy patient input OCT image, segmented retinal OCT image in light mode, and the corresponding CAM image for subject CO638. The three images provide a different view or transformation of the original OCT scan to illustrate the classification process and highlight regions of interest identified by the model. The grayscale image in figure \ref{CO638a1} shows the minimally processed OCT scan, depicting the layered structure of the retina. Figure \ref{CO638a2} appears to be a thresholded or binarized version of the original OCT image, where certain regions have been highlighted in white or left in grayscale. Figure \ref{CO638a3} is a CAM overlaid on the OCT scan, where colors indicate the areas of interest for the model's classification. In this healthy control, the retina's layers do not exhibit abnormal thickening or thinning, and the white areas seem to highlight non-retinal regions, reinforcing the model's correct classification. The CAM shows moderate attention across various retinal layers without an intense focus on any particular region, consistent with an absence of disease markers.


\begin{figure}[!ht]
    \centering
    \begin{subfigure}[b]{0.31\textwidth}
        \centering
         \includegraphics[width=\linewidth]{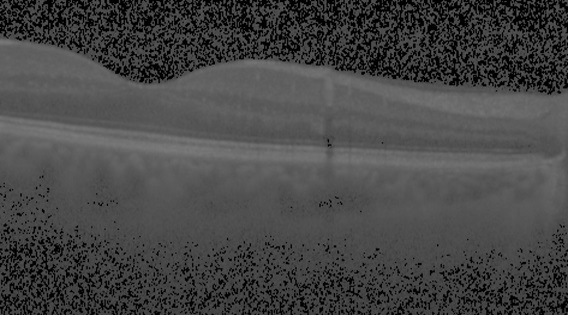}
        \caption{Input OCT image}
        \label{CO015a}
    \end{subfigure}
    \begin{subfigure}[b]{0.31\textwidth}
        \centering
         \includegraphics[width=\linewidth]{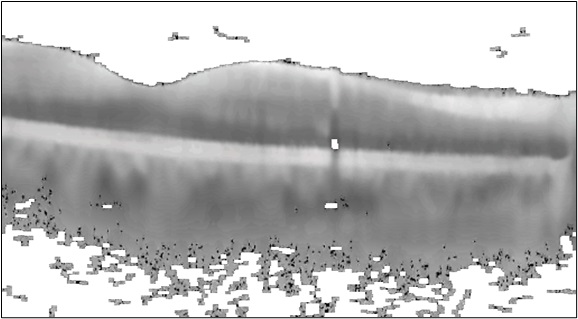}
        \caption{Segmented OCT image}
        \label{CO015b}
    \end{subfigure}
    \begin{subfigure}[b]{0.30\textwidth}
        \centering
        \includegraphics[width=\textwidth]{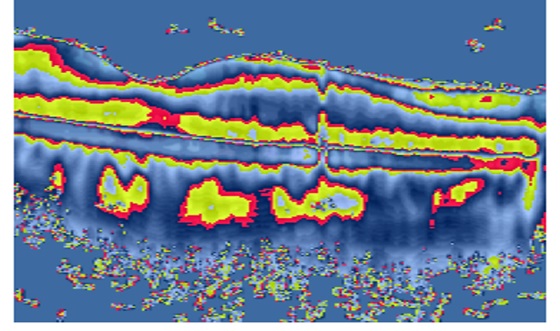}
        \caption{CAM OCT image}
        \label{CO015c}
    \end{subfigure}

    \caption{Correctly classified healthy controlled segmented OCT images of subject CO015}
    \label{fig:COSCC2}
\end{figure}

Figure \ref{fig:COSCC2} depicts the correctly classified healthy patient input OCT image, segmented retinal OCT image in light mode, and corresponding CAM image for subject CO015. This figure \ref{fig:COSCC2} illustrates a correctly classified healthy OCT image, where the model's focus is dispersed across the retinal layers without highlighting any area of abnormality. The model's balanced attention and the lack of disease markers in the OCT scan demonstrate the model's effectiveness in distinguishing between healthy and pathological OCT images.

\begin{figure}[!ht]
    \centering
    \begin{subfigure}[b]{0.29\textwidth}
        \centering
         \includegraphics[width=\linewidth]{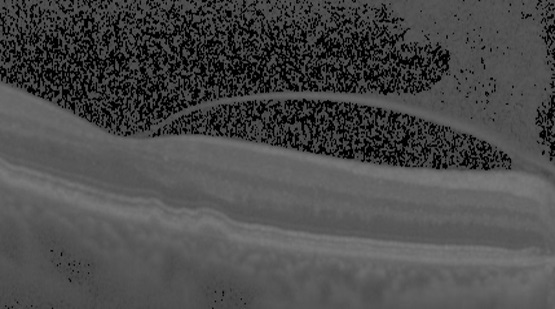}
        \caption{Input OCT image}
        \label{AD433a}
    \end{subfigure}
    \begin{subfigure}[b]{0.30\textwidth}
        \centering
         \includegraphics[width=\linewidth]{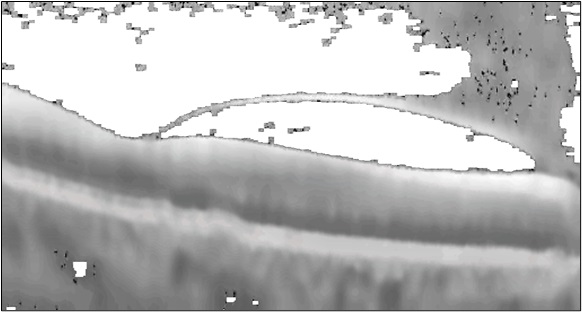}
        \caption{Segmented OCT image}
        \label{AD433b}
    \end{subfigure}
    \begin{subfigure}[b]{0.31\textwidth}
        \centering
        \includegraphics[width=\textwidth]{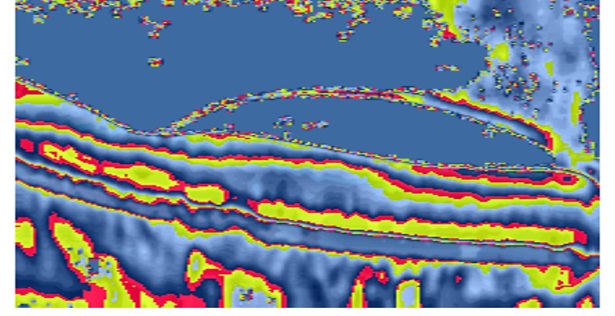}
        \caption{CAM OCT image}
        \label{AD433c}
    \end{subfigure}
    \caption{Misclassified Alzheimer's disease segmented OCT images of subject AD433.}
    \label{fig:MISSC1}
\end{figure}

Figure \ref{fig:MISSC1} depicts the misclassified AD patient segmented input retinal OCT image in light mode and the corresponding CAM image for subject AD433. This figure \ref{fig:MISSC1} shows three different representations of an OCT retinal scan image that has been misclassified. The CAM OCT image highlights the specific retinal layers and depicts activation regions less relevant to Alzheimer’s detection. 


\begin{figure}[!ht]
    \centering
    \begin{subfigure}[b]{0.29\textwidth}
        \centering
         \includegraphics[width=\linewidth]{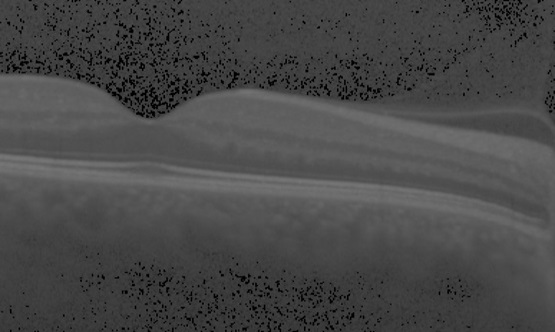}
        \caption{Input OCT image}
        \label{CO942a}
    \end{subfigure}
    \begin{subfigure}[b]{0.29\textwidth}
        \centering
         \includegraphics[width=\linewidth]{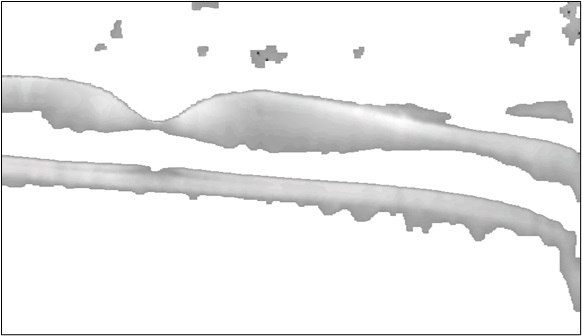}
        \caption{Segmented OCT image}
        \label{CO942b}
    \end{subfigure}
    \begin{subfigure}[b]{0.31\textwidth}
        \centering
        \includegraphics[width=\textwidth]{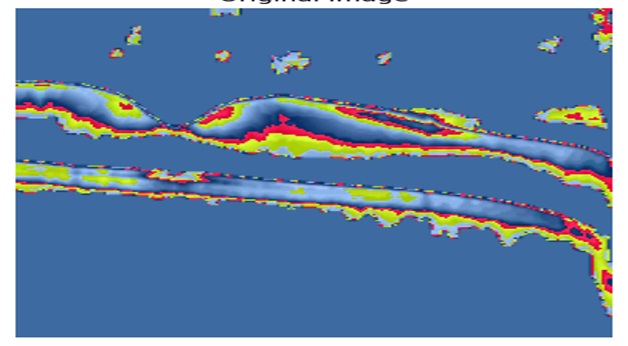}
        \caption{CAM OCT image}
        \label{CO942c}
    \end{subfigure}
    \caption{Misclassified healthy controlled segmented OCT image of subject CO942.}
    \label{fig:MISSC2}
\end{figure}

Figure \ref{fig:MISSC2} depicts the misclassified healthy controlled subject segmented input retinal OCT image in light mode and the corresponding CAM image for subject CO942. The CAM suggests that the model concentrates on areas outside the known retinal biomarkers for Alzheimer’s, such as the RNFL or GCL. Excessive focus on areas affected by noise or irrelevant structures leads to misclassification because these areas do not reliably indicate the disease. The binary thresholded image simplifies the retinal structure, potentially losing significant features that could contribute to misclassification.

\subsection{TransNetOCT}

The results highlight the TransNetOCT model's effectiveness in classifying OCT images, achieving an average accuracy of 98.18\% across 5-fold cross-validation. The precision of 0.9800, recall of 0.9846, and F1-score of 0.9815 demonstrate the model's ability to balance sensitivity and specificity, ensuring consistent and reliable classification. Additionally, the pixel-level accuracy of 98.18\% underscores its robustness in handling fine details in OCT images. The validation loss of 0.0471 and the MAE of 0.0182 confirm the model's precision and stability during training. The segmentation metrics, such as the Dice Similarity Coefficient (DSC) of 0.4945 and the infinite Hausdorff Distance (HD), are less relevant in this classification context. The metrics indicate reliable model predictions to improve segmentation consistency.

\begin{figure}[!ht]
    \centering
    \begin{subfigure}[b]{0.29\textwidth}
        \centering
         \includegraphics[width=1.0\linewidth, height=0.1\textheight]{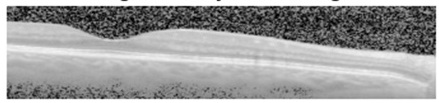}
        \caption{Input OCT image}
        \label{AD620a}
    \end{subfigure}
    \begin{subfigure}[b]{0.30\textwidth}
        \centering
         \includegraphics[width=1.0\linewidth, height=0.1\textheight]{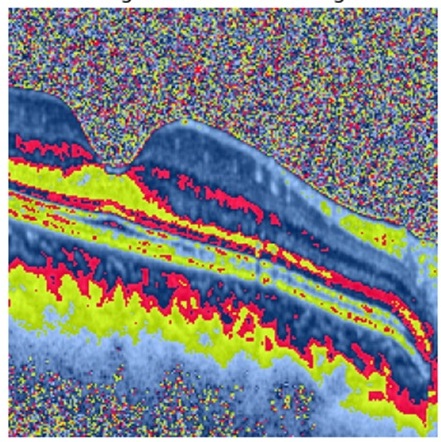}
        \caption{CAM OCT image}
        \label{AD620b}
    \end{subfigure}
    \begin{subfigure}[b]{0.31\textwidth}
        \centering
        \includegraphics[width=1.0\linewidth, height=0.1\textheight]{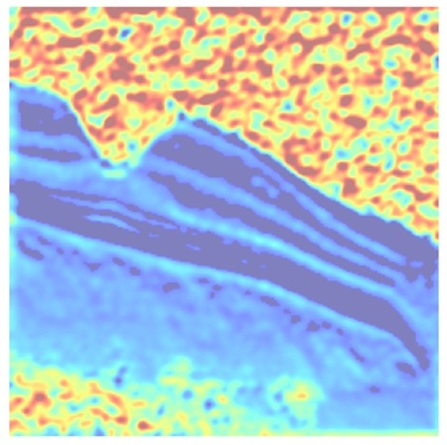}
        \caption{Filtered CAM OCT image}
        \label{AD620c}
    \end{subfigure}
    \caption{Correctly classified Alzheimer's disease OCT images of subject AD620.}
    \label{fig:TNOC1}
\end{figure}

Figure \ref{fig:TNOC1} shows a correctly classified AD retinal OCT scan (Predicted: AD, Actual: AD), demonstrating accurate model performance. The OCT grayscale image shows retinal layers consistent with an AD retina, while the CAM OCT image highlights regions with minor intensity variations that remain thinning of layers. The filtered CAM overlay indicates that the model focused on specific retinal boundaries and areas of structural consistency without overemphasizing non-relevant regions. This correct classification of the AD OCT image suggests that the model effectively identified the absence of pathological features, distinguishing natural variability from disease markers. 

\begin{figure}[!ht]
    \centering
    \begin{subfigure}[b]{0.29\textwidth}
        \centering
         \includegraphics[width=1.0\linewidth, height=0.1\textheight]{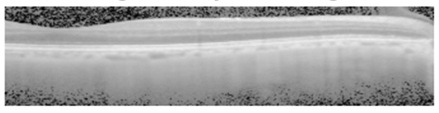}
        \caption{Input OCT image}
        \label{CO127a}
    \end{subfigure}
    \begin{subfigure}[b]{0.30\textwidth}
        \centering
         \includegraphics[width=1.0\linewidth, height=0.1\textheight]{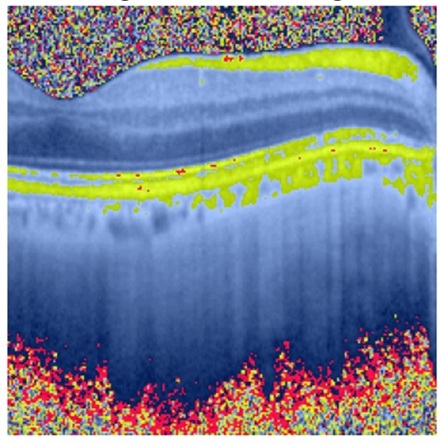}
        \caption{CAM OCT image}
        \label{CO127b}
    \end{subfigure}
    \begin{subfigure}[b]{0.31\textwidth}
        \centering
        \includegraphics[width=1.0\linewidth, height=0.1\textheight]{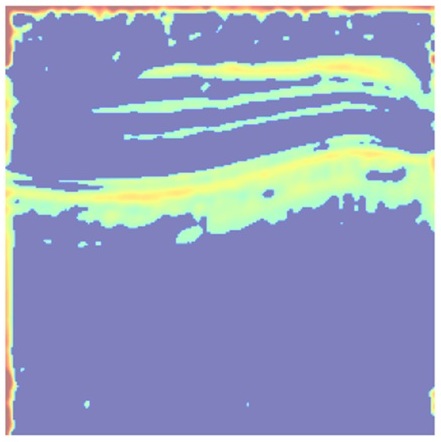}
        \caption{Filtered CAM OCT image}
        \label{CO127c}
    \end{subfigure}
    \caption{Correctly classified CO OCT images of subject CO127.}
    \label{fig:COtrans1}
\end{figure}

Figure \ref{fig:COtrans1} shows a correctly classified CO retinal OCT scan (Predicted: CO, Actual: CO). The grayscale OCT image shows retinal layers consistent with a healthy retina, while the CAM OCT image highlights regions with minor intensity variations that remain within normal limits. The filtered CAM overlay indicates that the model focused on specific retinal boundaries such as RNFL and choroid layer, which are biomarkers of AD OCT image.

\begin{figure}[!ht]
    \centering
    \begin{subfigure}[b]{0.29\textwidth}
        \centering
         \includegraphics[width=1.0\linewidth, height=0.1\textheight]{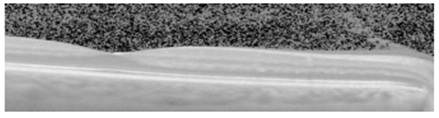}
        \caption{Input OCT image}
        \label{AD613a}
    \end{subfigure}
    \begin{subfigure}[b]{0.30\textwidth}
        \centering
         \includegraphics[width=1.0\linewidth, height=0.1\textheight]{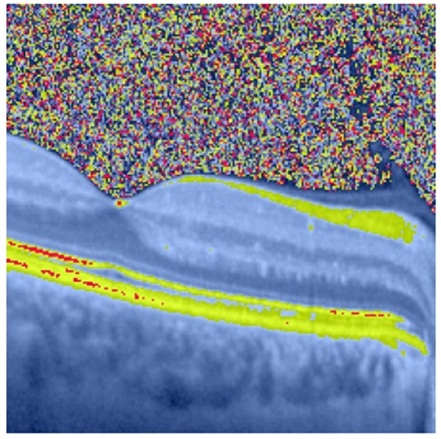}
        \caption{CAM OCT image}
        \label{AD613b}
    \end{subfigure}
    \begin{subfigure}[b]{0.31\textwidth}
        \centering
        \includegraphics[width=1.0\linewidth, height=0.1\textheight]{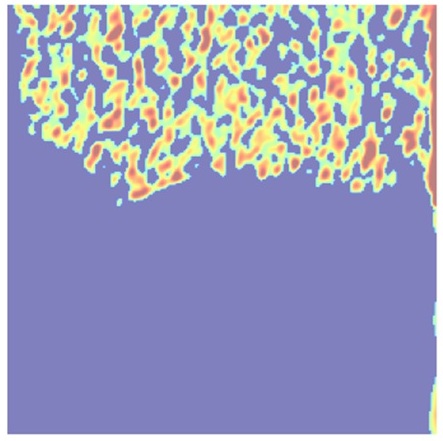}
        \caption{Filtered CAM OCT image}
        \label{AD613}
    \end{subfigure}
    \caption{Misclassified Alzheimer's disease OCT images of subject AD613.}
    \label{fig:ADmistrans1}
\end{figure}

Figure \ref{fig:ADmistrans1} showcases a misclassified retinal OCT scan from an Alzheimer's disease study, highlighting a predicted label of CO versus the actual label of AD. The visualization includes the original grayscale OCT image, a CAM image emphasizing key retinal structures, and a filtered CAM to pinpoint RNFL and choroid regions influencing the model's prediction. Despite the intricate visualization, critical features in the retinal layers contributing to Alzheimer's classification have been likely misinterpreted by the model in Alzheimer's diagnosis using OCT scans.

\begin{figure}[!ht]
    \centering
    \begin{subfigure}[b]{0.29\textwidth}
        \centering
         \includegraphics[width=1.0\linewidth, height=0.1\textheight]{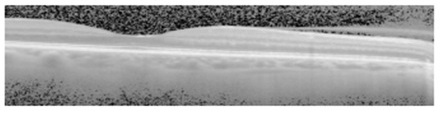}
        \caption{Input OCT image}
        \label{CO443a}
    \end{subfigure}
    \begin{subfigure}[b]{0.30\textwidth}
        \centering
         \includegraphics[width=1.0\linewidth, height=0.1\textheight]{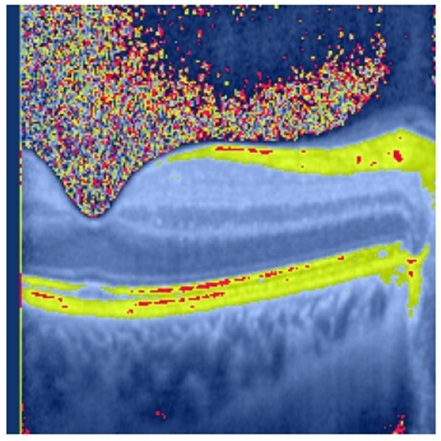}
        \caption{CAM OCT image}
        \label{CO443b}
    \end{subfigure}
    \begin{subfigure}[b]{0.31\textwidth}
        \centering
        \includegraphics[width=1.0\linewidth, height=0.1\textheight]{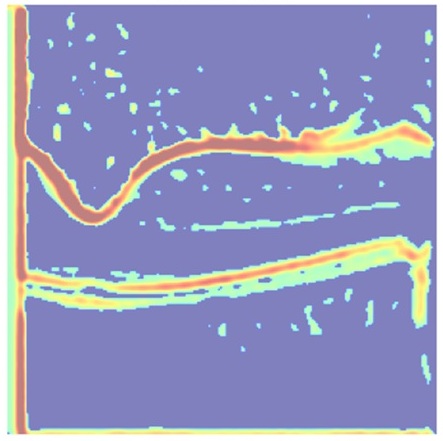}
        \caption{Filtered CAM OCT image}
        \label{CO443c}
    \end{subfigure}
    \caption{Misclassified CO OCT images of subject CO443.}
    \label{fig:COmistrans1}
\end{figure}

Figure \ref{fig:COmistrans1} represents a healthy control retinal OCT scan that the model misclassified, which has been predicted as AD but actually CO. The input OCT image shows smooth and continuous retinal layers without visible abnormalities. However, the filtered CAM reveals that the model incorrectly focused on RNFL and choroid layers, misinterpreting natural structural variations as indicators of disease. 


The results of the TransnetOCT model evaluation indicate the classification of segmented OCT images with an average accuracy of 98.91\% on 5-fold cross-validation, with a precision of 0.9867, a recall of 0.9926 and an F1 score of 0.9893, reflecting a well-balanced classification ability. The segmentation metrics reveal an average ASSD (Average Symmetric Surface Distance) of 0.011, demonstrating a high spatial alignment accuracy. However, the average Dice Similarity Coefficient (DSC) of 0.4982 suggests an improvement in overlap consistency. Additional performance indicators include a validation loss of 0.0355, an MAE of 0.0109, and an impressive pixel-level accuracy of 98.91\%.

\begin{figure}[!ht]
    \centering
    \begin{subfigure}[b]{0.29\textwidth}
        \centering
         \includegraphics[width=1.0\linewidth, height=0.1\textheight]{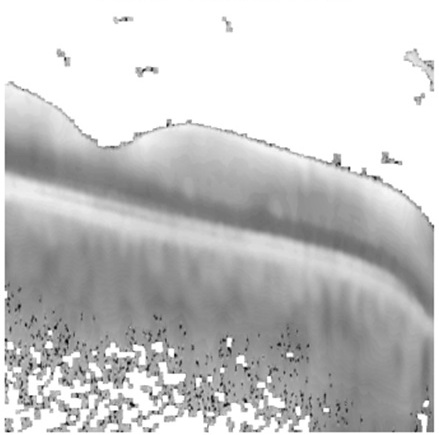}
        \caption{Input segmented OCT image}
        \label{AD508a}
    \end{subfigure}
    \begin{subfigure}[b]{0.30\textwidth}
        \centering
         \includegraphics[width=1.0\linewidth, height=0.1\textheight]{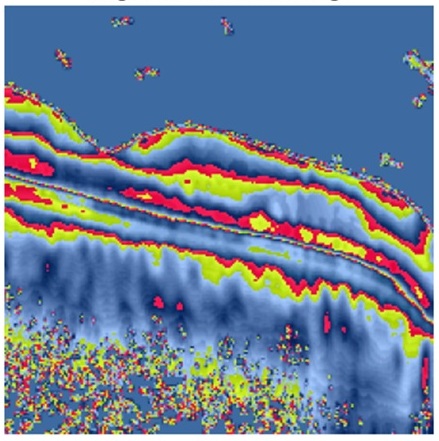}
        \caption{CAM OCT image}
        \label{AD508b}
    \end{subfigure}
    \begin{subfigure}[b]{0.31\textwidth}
        \centering
        \includegraphics[width=1.0\linewidth, height=0.1\textheight]{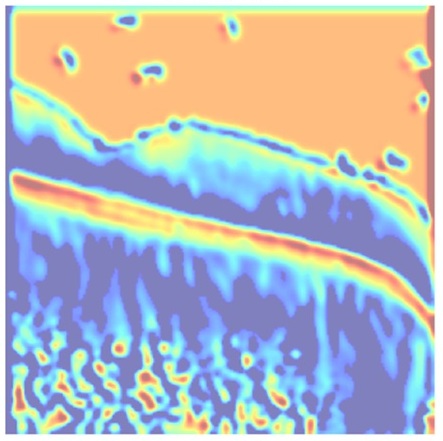}
        \caption{Filtered CAM OCT image}
        \label{AD508c}
    \end{subfigure}
    \caption{Correctly classified Alzheimer's disease segmented OCT images of subject AD508.}
    \label{fig:ADsegtrans1}
\end{figure}

Figure \ref{fig:ADsegtrans1} shows an OCT scan of the retina correctly classified as representing a subject with Alzheimer's disease (predicted label: AD, actual label: CO). It includes three panels: the input segmented image, the CAM image emphasizing details of the retinal layer, and the filtered CAM overlay highlighting critical regions influencing the classification. The filtered CAM  indicates that the model focused on specific structural regions in the retina, such as boundary layers and central thickness, which likely indicate Alzheimer' s-related retinal changes. The consistent focus on these significant features validates the model's accuracy in distinguishing Alzheimer's patterns, demonstrating its capability to extract disease-relevant biomarkers from retinal OCT scans.

\begin{figure}[!ht]
    \centering
    \begin{subfigure}[b]{0.29\textwidth}
        \centering
         \includegraphics[width=1.0\linewidth, height=0.1\textheight]{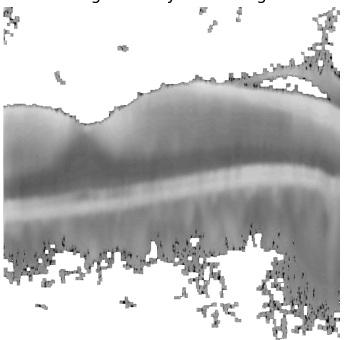}
        \caption{Input segmented OCT image}
        \label{CO091a}
    \end{subfigure}
    \begin{subfigure}[b]{0.30\textwidth}
        \centering
         \includegraphics[width=1.0\linewidth, height=0.1\textheight]{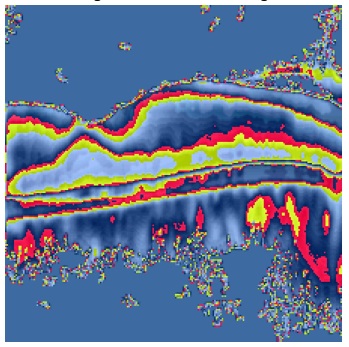}
        \caption{CAM OCT image}
        \label{CO091b}
    \end{subfigure}
    \begin{subfigure}[b]{0.31\textwidth}
        \centering
        \includegraphics[width=1.0\linewidth, height=0.1\textheight]{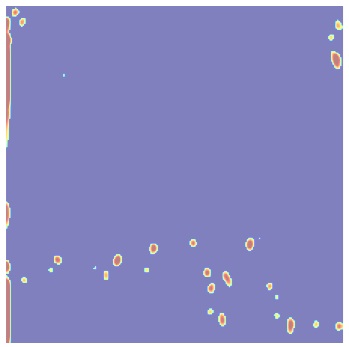}
        \caption{Filtered CAM OCT image}
        \label{CO091c}
    \end{subfigure}
    \caption{Correctly classified CO segmented OCT images of subject CO091.}
     \label{fig:COsegtrans1}
\end{figure}

Figure \ref{fig:COsegtrans1} shows a correctly classified CO retinal OCT (predicted: CO, actual: CO), demonstrating the model's ability to accurately identify structural features associated with a healthy retina. The input-segmented OCT image highlights these layers with clear boundary delineations and normal thickness patterns. The filtered CAM demonstrates that the model correctly focused on the segmented retinal layers, emphasizing regions of interest without misinterpreting natural variations as abnormalities. This correct classification indicates that the model effectively utilizes segmentation to analyze layer-specific features, reinforcing its ability to differentiate between healthy and pathological scans based on structural integrity and consistent patterns in the retinal OCT image.

\begin{figure}[!ht]
    \centering
    \begin{subfigure}[b]{0.29\textwidth}
        \centering
         \includegraphics[width=1.0\linewidth, height=0.09\textheight]{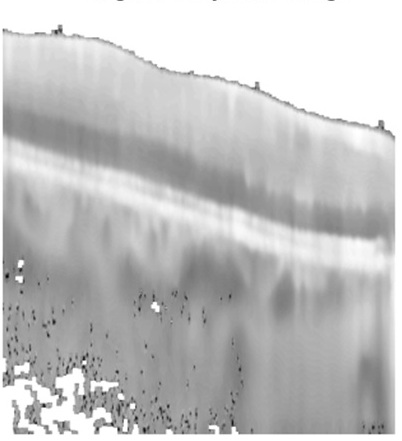}
        \caption{Input segmented OCT image}
        \label{AD727a}
    \end{subfigure}
    \begin{subfigure}[b]{0.30\textwidth}
        \centering
         \includegraphics[width=1.0\linewidth, height=0.1\textheight]{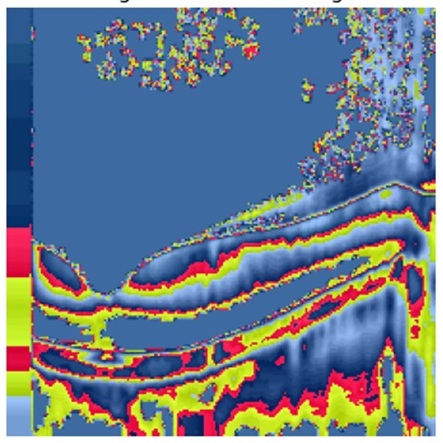}
        \caption{CAM OCT image}
        \label{AD727b}
    \end{subfigure}
    \begin{subfigure}[b]{0.31\textwidth}
        \centering
        \includegraphics[width=1.0\linewidth, height=0.1\textheight]{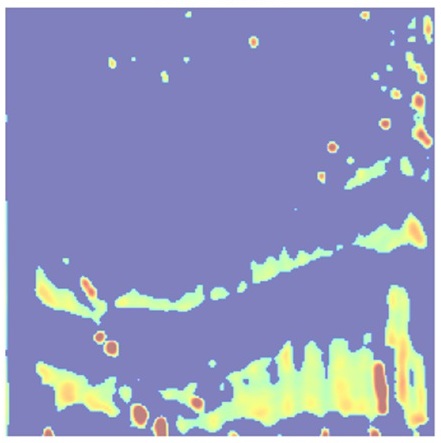}
        \caption{Filtered CAM OCT image}
        \label{AD727c}
    \end{subfigure}
    \caption{Misclassified Alzheimer's disease segmented  OCT images of subject AD727.}
    \label{fig:ADmissegtrans1}
\end{figure}

Figure \ref{fig:ADmissegtrans1} showcases a misclassified retinal segmented OCT scan from an Alzheimer's disease study, highlighting a predicted label of CO versus the actual label of AD. The visualization includes the input OCT image, a CAM image emphasizing key retinal structures, and a filtered CAM to pinpoint RNFL AND choroid regions influencing the model's prediction. 


\begin{figure}[!ht]
    \centering
    \begin{subfigure}[b]{0.29\textwidth}
        \centering
         \includegraphics[width=1.0\linewidth, height=0.12\textheight]{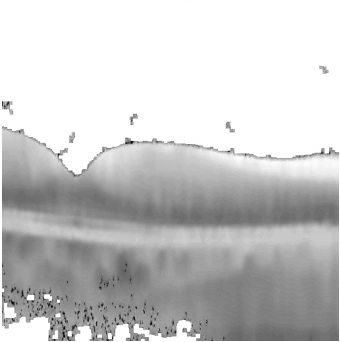}
        \caption{Input segmented OCT image}
        \label{CO638a}
    \end{subfigure}
    \begin{subfigure}[b]{0.30\textwidth}
        \centering
         \includegraphics[width=1.0\linewidth, height=0.1\textheight]{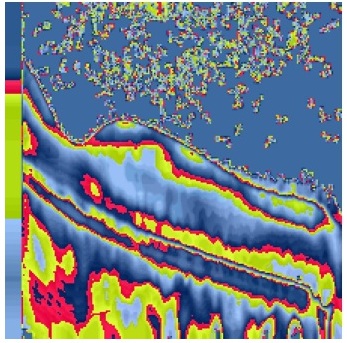}
        \caption{CAM OCT image}
        \label{CO638b}
    \end{subfigure}
    \begin{subfigure}[b]{0.31\textwidth}
        \centering
        \includegraphics[width=1.0\linewidth, height=0.1\textheight]{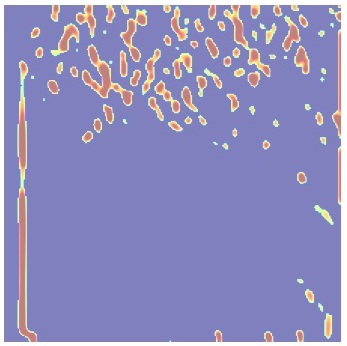}
        \caption{Filtered CAM OCT image}
        \label{CO638c}
    \end{subfigure}
    \caption{Misclassified CO segmented OCT images of subject CO638.}
    \label{fig:COmissegtrans1}
\end{figure}

Figure \ref{fig:COmissegtrans1} represents a CO retinal segmented OCT scan that the model misclassified (Predicted: AD, Actual: CO). The input OCT image shows smooth and continuous retinal layers without visible abnormalities, characteristic of a healthy retina. 



\subsection{Discussion}

The original OCT image dataset gives a better accuracy of 93.45\% than the segmented OCT image data set, which offers an accuracy of 89. 82\% using the Swin transformer model. It is observed that information is lost while segmenting the OCT images. While both datasets yield high accuracies, the model performs slightly better on unsegmented images, particularly evident in the first table's consistent accuracy above 90\% across all folds, including one fold reaching 100\%. In contrast, the segmented OCT dataset in the second table shows more variability, with accuracy dropping to 83.64\%  in the fifth fold and increasing misclassifications as indicated by higher counts of false positives and false negatives in specific folds. This suggests that the Swin Transformer may be more effective when trained on unsegmented OCT images, possibly due to more contextual information that aids classification. 

The Swin transformer model application in medical image classification effectively differentiates Alzheimer's disease and healthy controls. The findings suggest that advanced deep-learning models have significantly enhanced diagnostic accuracy and efficiency in neurodegenerative diseases.

The evaluation results across the TransnetOCT model demonstrate exceptional performance in OCT image classification, with high average accuracies of 98.18\% for original OCT images and 98.91 for segmented OCT images with balanced precision, recall, and F1-scores. These metrics highlight the models' robustness and reliability in distinguishing minor preprocessed and segmented OCT images. The strong pixel-level accuracies (0.9818 and 0.9891) and low validation losses (0.0471 and 0.0355) underscore the stability and precision achieved during training. Despite these strengths, segmentation-related metrics such as the Dice Similarity Coefficient (0.4945 and 0.4982) and infinite Hausdorff Distance reveal limitations in overlap consistency and sensitivity to outliers. The ASSD of 0.011 in the segmented dataset demonstrates its spatial alignment accuracy, reflecting its potential to handle structural variations.

This study tried to identify critical retinal OCT regions contributing to the classification of Alzheimer's disease and healthy subjects using advanced interpretability techniques, including Grad-CAM, saliency maps, integrated gradients, and occlusion analysis for non-segmented and segmented OCT images. These methods explored the patterns and features within the OCT images. Despite leveraging these state-of-the-art approaches, the Swin transformer model's limitation is that it cannot differentiate distinct and consistent regions that reliably distinguish between the two classes. However, the TransNetOCT model has been achieved to differentiate AD and CO. This outcome highlights the complexity of the underlying thin layers of retinal OCT images and also due to the limited availability of public datasets with few subjects. It suggests for future work that additional factors, such as higher-resolution data, multimodal inputs, or alternative model architectures, are necessary to achieve more robust classification insights.

Many non-neurologic diseases can have retinal manifestations indistinguishable from the features reportedly used by current machine learning models to distinguish between healthy eyes and those with neurodegenerative diseases. Future machine learning models will need to use more diverse datasets based on longitudinal data to evaluate whether they can identify specific features that differentiate actual neurodegenerative diseases from other diseases with neuroretinal implications.

\section{Conclusion}

The advancement of Alzheimer's disease classification using retinal OCT images reveals essential insights into the productiveness of various deep-learning architectures. Among the eight architectures compared, DenseNet121 Swin Transformer, and TransNetOCT stand out with the highest accuracy rates of 76.36\%, 93.45\% and 98.18\%, respectively. These classification results emphasize the primacy of advanced deep learning pre-trained models, particularly transformer models, in capturing the intricate patterns corresponding with Alzheimer's disease in retinal OCT images. The CNN models such as Simple CNN, VGG16, and RESNET50 demonstrate moderate to low performance, highlighting their limitations in this OCT image classification context. The findings suggest that leveraging modern architectures, especially those based on transformer frameworks, can enhance classification accuracy, ultimately aiding in more reliable and timely diagnoses of Alzheimer's disease through retinal OCT imaging.

The CAM with five-fold cross-validation allows for both interpretability and reliability in the TransNetOCT model's performance. The CAM visualizations ensure the model focuses on appropriate image regions, such as RFNL and the choroid layer. At the same time, cross-validation provides confidence that the model is generalized effectively to unseen data. These results underscore the Swin Transformer's potential for tasks requiring high accuracy and transparency in classifying Alzheimer's disease OCT images from healthy controlled OCT images.

 This proposed work highlights the potential for further research and application of transformer models in improving diagnostic tools for neurodegenerative conditions. Expanding the work for the larger dataset and exploring additional architectures to enhance model performance further will be extended in future work.


\section*{Declarations}

\begin{itemize}
\item Ethics approval This paper does not contain any experiments with
human participants or animals performed by any of the authors.
\item Conflict of interest The authors declare no competing interests.
\end{itemize}

\bibliographystyle{unsrt}  

\bibliography{references}

\end{document}